\documentclass[9pt,conference]{IEEEtran}
\IEEEoverridecommandlockouts

\usepackage{booktabs}  

\usepackage{graphicx}
\usepackage{amsmath,amsthm}
\usepackage{amssymb}
\usepackage{color}
\usepackage{url}
\usepackage{multirow}
\usepackage[linesnumbered,ruled]{algorithm2e}
\usepackage{verbatim}
\usepackage{subfigure}
\usepackage{braket}
\usepackage{tikz}
\usetikzlibrary{quantikz}
\usepackage{flowchart}
\usepackage[numbers,sort&compress]{natbib}

\usepackage{multirow}
\usepackage{threeparttable}

\newcommand{\tr}{{\rm tr}}
\newcommand{\e}{\mathcal{E}}
\newcommand{\n}{\mathcal{N}}

\usepackage{xcolor}

\newtheorem{defn}{Definition}

\newtheorem{exam}{Example}

\newtheorem{prob}{Problem}
\usepackage{authblk}

\title{Approximate Equivalence Checking of Noisy Quantum Circuits}

\author[1]{Xin~Hong}
\author[1,2,3*]{Mingsheng~Ying\thanks{*Corresponding authors:\{mingsheng.ying,yuan.feng,sanjiang.li\}@uts.edu.au.}}
\author[1*]{Yuan~Feng}
\author[1,4]{Xiangzhen~Zhou}
\author[1*]{Sanjiang~Li}
\affil[1]{Centre for Quantum Software and Information, University of Technology Sydney, Australia}
\affil[2]{State Key Laboratory of Computer Science, Institute of Software, Chinese Academy of Sciences, China}
\affil[3]{Department of Computer Science and Technology, Tsinghua University, China}
\affil[4]{State Key Lab of Millimeter Waves, Southeast University, China \vspace{-0.5em}}

\makeatletter
\def\ps@IEEEtitlepagestyle{%
  \def\@oddfoot{\mycopyrightnotice}%
  \def\@evenfoot{}%
}
\def\mycopyrightnotice{%
  {\large 978-1-6654-3274-0/21/\$31.00~\copyright2021 IEEE \hfill}
}
\makeatother

\begin{document}

\maketitle
\begin{abstract}

We study the fundamental design automation problem of equivalence checking in the NISQ (Noisy Intermediate-Scale Quantum) computing realm where quantum noise is present inevitably. The notion of approximate equivalence of (possibly noisy) quantum circuits is defined based on the Jamiolkowski fidelity which measures the average distance between output states of two super-operators when the input is chosen at random. By employing tensor network contraction, we present two algorithms, aiming at different situations where the number of noises varies, for computing the fidelity between an ideal quantum circuit and its noisy implementation. The effectiveness of our algorithms is demonstrated by experimenting on benchmarks of real NISQ circuits. When compared with the state-of-the-art implementation incorporated in Qiskit, 
experimental results show that the proposed algorithms outperform in both efficiency and scalability.

\end{abstract}


\section{Introduction}

Equivalence checking techniques have been widely employed in the EDA (Electronic Design Automation) industry to check whether two different circuit designs exhibit exactly the same behaviour. This is very important to ensure that a design is error-free. Important techniques such as BDD \cite{bryant1986graph,molitor2007equivalence} and SAT \cite{mishchenko2006improvements,molitor2007equivalence} have been successful in such a task in the classical case.

Nowadays, quantum computing has attracted great attention from both industry and academia due to its potential speedup in solving problems such as integer factorisation \cite{shor1999polynomial}, database search \cite{grover1996fast}, and many machine learning tasks \cite{biamonte2017quantum}. Experiments have been proposed to demonstrate quantum supremacy \cite{GoogleQsupr} and to explore the possibilities of simulating a chemical system \cite{cao2019quantum}. International leading IT companies such as IBM and Google 
have invested enormous resources to develop quantum hardware and software. It is expected that quantum devices with several hundreds valid qubits will appear very soon.

As quantum circuits become larger and larger, they are more and more error-prone. Equivalence checking of quantum circuits has been discussed in \cite{Via07, Markov, burgholzer2020advanced,wille2009equivalence,burgholzer2020power,wille2013compact}, where it is assumed that 
each gate in either circuit is represented as a unitary operation and the aim is to check if the two unitary matrices representing the two circuits are equal (up to a global phase). Often, canonical decision diagram representations for the two circuits are constructed and the two circuits are equivalent if and only if they have the same representation.

On the other hand, quantum computers we have at present or in the near future are all Noisy Intermediate-Scale Quantum (NISQ) devices and quantum circuits executed on them are all plagued by more or less noise. For such kinds of circuits, the proposed approaches for equivalence checking would fail as noisy gates cannot be represented by unitary matrices, and a qualitative answer (being equivalent or not) does not make sense. 
Thus, 
it is necessary and very important to develop new methods for checking the approximate equivalence, or computing the distance, of noisy circuits.



This paper aims at proposing a proper definition as well as efficient algorithms for the problem of  approximate equivalence between one ideal (unitary) circuit and its noisy implementation.
The key notion is the Jamiolkowski fidelity~\cite{raginsky2001fidelity} which measures the average distance between output states of two super-operators when the input is chosen at random. In addition to the clear physical interpretation, Jamiolkowski fidelity enjoys some nice properties such as stability and chaining~\cite{Gil05} which guarantee that the error scales at most linearly when smaller noisy quantum circuits are concatenated into a larger one. 
More importantly, computing this measure of distance can be reduced to the calculation of traces of some (non-unitary) matrices obtained from the original circuits, a task which can be done efficiently 
by employing tensor network contraction.
With these observations, we present two algorithms,
aiming at different situations where the number of noises varies, and implement them by using a recently developed software package --- Tensor Decision Diagram (TDD)~\cite{tdd} which
represents a tensor as a decision diagram and supports efficient contraction of tensor networks.
The effectiveness of our algorithms and implementation is demonstrated by experimenting on 
real NISQ circuits. Experimental results show that 
our algorithms outperform in both efficiency and scalability over the state-of-the-art implementation incorporated in Qiskit \cite{aleksandrowicz2019qiskit}.

{\bf Related works.} 
Existing works in equivalence checking of quantum circuits are based on Decision Diagrams (DDs) or SAT or a combination of them~\cite{Via07, Markov, burgholzer2020advanced,wille2009equivalence,burgholzer2020power,wille2013compact}. In particular, QMDD \cite{niemann2015qmdds}, which provides a more economical way to represent unitary operators when compared with array-based representations, is used in \cite{burgholzer2020advanced} as the underlying data structure for equivalence checking. To our best knowledge, there are no attempts in extending these methods to deal with approximate equivalence of noisy quantum circuits.




Distance measures of quantum processes have been considered in \cite{Gil05,belavkin2005operational}. In particular, the Jamiolkowski fidelity is identified in \cite{Gil05} as one of the best measures for comparing quantum processes. Biamonte \cite{biamonte2019lectures} describes a method for calculating the average gate fidelity based on tensor networks, where quantum processes are expressed in terms of Choi-matrices. However, these matrices are not convenient to compute from the classical description of quantum noisy circuits.

The important notion of reversible miter is introduced in \cite{Markov} for equivalence checking of (unitary) quantum circuits. For two circuits $C, D$, a miter is constructed by concatenating $C$ with the reverse of $D$, and the two circuits are equivalent if and only if the miter implements the identity operator. When combined with local optimisations like cancelling two consecutive gates, this technique can significantly simplify the circuits and thus the process of equivalence checking.  Such a miter is also used in \cite{burgholzer2020advanced}, where a more efficient method for equivalence checking is designed by carefully choosing the gates to be calculated every time. Our algorithms also rely on calculating the traces of similar miter constructions.




Classical simulation of noisy quantum circuits are considered in \cite{gao2018efficient} and \cite{grurl2020considering}. 
Gao and Duan \cite{gao2018efficient} develop a tensor network 
tool to represent the ensemble of noisy quantum circuits, while Grurl et al. \cite{grurl2020considering} represent density matrices as decision diagrams and show that considering decoherence errors does not necessarily affect the compactness of the decision diagram representations. Li et al. \cite{li2020eliminating} propose an optimisation technology for the Monte Carlo simulation of noisy quantum circuits, where they exploit the structure similarity of an ensemble of noisy circuits to be simulated and try to maximally reuse precomputed states. Such a technology is inherently used by decision diagrams, where a single shared Hash table, called computed table, is used to store all nodes of the decision diagrams generated in the whole calculation process.

\section{Quantum Circuits and Noise}\label{Sec_QCur}
In this section, we review some basic concepts from quantum computing to help understand noisy quantum circuits.

\subsection{Quantum Circuits}

In classical digital computation, data are  represented by a string of bits. When sending through a classical circuit, the state of the input will be transformed by a sequence of logical gates. In quantum computing, the counterpart of bit is called \emph{qubit}. The state of a qubit is often represented in  Dirac notation
$
	\ket{ \psi} =  \alpha_0 \ket{0} + \alpha_1 \ket{1},
$
where $\alpha_0$ and $\alpha_1$ are complex numbers, called the amplitudes of $\ket{\psi}$, and ${\left| \alpha_0  \right|^2} + {\left| \alpha_1  \right|^2} = 1$. We also use the vector $[\alpha_0,\alpha_1]^T$ to represent a single-qubit state. In general, an $n$-qubit quantum state is represented as a $2^n$-dimensional complex vector  $[\alpha_0,\alpha_1,\dots,\alpha_{2^n-1}]^T$.

The evolution of a quantum system is described by a unitary transformation, which is usually called a \emph{quantum gate} in quantum computing.
Generally, an $n$-qubit quantum gate is represented as a $2^n\times 2^n$-dimensional unitary transformation matrix.
Typical 1-qubit gates used in this paper include Pauli gates
{\small
$$X=\left[\begin{array}{cc} 0 & 1\\ 1 & 0\end{array}\right],\ Y=\left[\begin{array}{cc} 0 & -i\\  i & 0\end{array}\right],\ Z=\left[\begin{array}{cc} 1 & 0 \\ 0 & -1\end{array}\right],$$}
\hspace{-.8em} Hadamard gate $H= (X+Z)/\sqrt{2}$, and phase gate $S = \sqrt{Z}$.
The quantum state after applying a gate can be obtained by multiplying the corresponding unitary matrix and the vector that represents the input state. 
For example, the output state resulted from applying a Hadamard gate to $\ket{ \psi}$ defined above is calculated as follows
{\small 
\[\frac{1}{{\sqrt 2 }}\left[ {\begin{array}{*{20}{r}}
		1&1\\
		1&{ - 1}
\end{array}} \right]\left[ {\begin{array}{*{20}{c}}
		{{\alpha _0}}\\
		{{\alpha _1}}
\end{array}} \right] = \frac{1}{{\sqrt 2 }}\left[ {\begin{array}{*{20}{c}}
		{{\alpha _0} + {\alpha _1}}\\
		{{\alpha _0} - {\alpha _1}}
\end{array}} \right].\]
}

A \emph{quantum circuit} consists of a set of qubits and a sequence of quantum gates. Given an input state, quantum gates in the circuit are applied to the corresponding qubits in a sequential manner. 
Clearly, an $n$-qubit quantum circuit also describes a functionality represented as a $2^n\times 2^n$-dimensional unitary matrix.

\begin{figure}
\scalebox{0.8}{
\centerline{
\begin{quantikz}
\lstick{$q_1$}  & \gate{H}    &\gate{S}   &    \qw           &\swap{1}  &\qw\\
\lstick{$q_2$}  & \qw           &\ctrl{-1}    &     \gate{H}  &\targX{}    &\qw\\
\end{quantikz}
}
}
\caption{Circuit for 2-qubit quantum Fourier transform (QFT).}
\label{exp-for-quantum-circuit}
\end{figure}
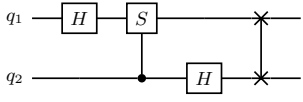

\begin{exam}\label{ex-q-circuit}
Depicted in Fig.~\ref{exp-for-quantum-circuit} is a circuit which implements the 2-qubit quantum Fourier transform (QFT)~\cite{NC00}, where \\

\scalebox{0.7}{
\centerline{$\quad\quad\quad\quad
	\begin{quantikz}
 	\gate{S}\\
	\ctrl{-1}\\
	\end{quantikz} 
	=\left[\begin{array}{cccc} 
		1 & 0 & 0 & 0\\ 
		0 & 1 & 0 & 0\\
		0 & 0 & 1 & 0\\
		0 & 0 & 0 & i
	\end{array}\right],\quad\quad 
	\begin{quantikz}
	\swap{1}\\
	\targX{} 
	\end{quantikz}
	= \mathit{SWAP}  =\left[\begin{array}{cccc} 
		1 & 0 & 0 & 0\\ 
		0 & 0 & 1 & 0\\
		0 & 1 & 0 & 0\\
		0 & 0 & 0 & 1
\end{array}\right].
$}}
\end{exam}
For simplicity, we often use the same symbol, say $U$, to denote both a quantum gate (or circuit) and its corresponding unitary matrix. 

\subsection{Noisy quantum circuits}

The pure state-unitary transformation framework presented above 
works perfectly well for ideal (noiseless) quantum computing.
However, to verify noisy implementation of ideal quantum circuits,  which is the main purpose of the current paper, we have to extend this framework to accommodate mixed states and  super-operators.

Mathematically, a \emph{mixed state} of an $n$-qubit system is described by a $2^n\times 2^n$ density matrix, i.e., a positive semi-definite matrix $\rho$ with $\tr(\rho)=1$. Obviously, for a pure state $|\psi\rangle$, the outer product $|\psi\rangle\langle\psi|$, denoted by $\psi$ henceforth, of $|\psi\rangle$ with its complex conjugate $\bra{\psi}$ is a mixed state. 

Dynamics of an $n$-qubit system whose states are given as mixed states is modelled by a \emph{super-operator} which is a linear map $\mathcal{E}$ between $2^n\times 2^n$ density matrices. A convenient way of representing an $n$-qubit super-operator $\e$ is the \textit{Kraus operator-sum form}: there exist a set of $2^n\times 2^n$ matrices $\{E_i\}$ satisfying the normalisation condition $\sum_{i}E_i^{\dag}E_i= I_{2^n}$, such that for each mixed state $\rho$,  $\mathcal{E}(\rho)=\sum_{i}E_i\rho E_i^{\dag}.$ In this case, we write $\mathcal{E}=\left\{E_i\right\}$. Note that unitary operator $U$ can also be seen as a super-operator $\mathcal{U}:\rho\mapsto U\rho U^\dag$. 


Noises in quantum computing can be naturally represented as super-operators (in the Kraus operator-sum form).
\begin{exam}\label{flips}  Given $p\in [0,1]$,
	several canonical noises on a single qubit are: 
\begin{enumerate}\item Bit flip. This kind of noise flips the state of a qubit from $|0\rangle$ to $|1\rangle$ and vice versa with probability $1-p$. It is modelled by the super-operator $\mathcal{N}_{\mathit{bf}}(\rho)=p\rho+(1-p)X\rho X$.  
\item Phase flip. This noise applies a phase operator $Z$ on the target qubit with probability $1-p$. It is given as the super-operator $\mathcal{N}_{\mathit{pf}}(\rho)=p\rho+(1-p)Z\rho Z$.  
\item Bit-phase flip. This noise applies $Y$ on the target qubit with probability $1-p$: $\mathcal{N}_{\mathit{bpf}}(\rho)=p\rho+(1-p)Y\rho Y$. Obviously, it is a combination of a bit-flip and a phase flip because $Y=iXZ$.
\item Depolarisation. The depolarisation of a qubit with parameter $p$ is modelled by $\mathcal{N}_{dep}(\rho)=p\rho+\frac{1-p}{3}(X\rho X+Y\rho Y+Z\rho Z)$.
\end{enumerate}\end{exam}

\begin{figure}
\centering
\scalebox{0.8}{
\begin{quantikz}
	\lstick{$q_1$}  & \gate{H}     &\gate{S}   &    \gate{\mathcal{N}'}          &\swap{1}  &\qw\\
	\lstick{$q_2$}  & \gate{\mathcal{N}}         &\ctrl{-1}    &     \gate{H}  &\targX{}    &\qw\\
\end{quantikz}
}
\caption{A noisy circuit for 2-qubit QFT with two noises inserted.}
    \label{exp-for-noisy-circuit}
\end{figure}
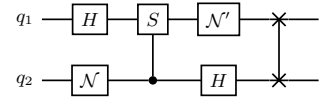

Now by allowing occurrence of noises, we can easily extend the notions of quantum gates and circuits to the noisy ones. Specifically,
a noisy quantum circuit is composed of noisy quantum gates which are represented as super-operators instead of only unitary operators.
Since unitary operators can be regarded as super-operators, noiseless gates and circuits are special cases of their noisy counterparts. As a simple example, Fig.~\ref{exp-for-noisy-circuit} shows a noisy implementation of the 2-qubit QFT 
where two 1-qubit noises $\n$ and $\n'$ are introduced.

Again, we abuse the notation slightly to use the same symbol to denote both a noisy quantum gate (or circuit) and its corresponding super-operator. However, to avoid confusion, we usually use $\n$, $\n'$, etc., for noisy gates and $\e$, $\e'$, etc., for noisy circuits.

\section{Approximate Equivalence of Quantum Circuits}\label{Sec_approxi_equi}


This section aims at formalising the problem of approximate equivalence checking of noisy quantum circuits. The key notion is an appropriate definition of distance between the ideal circuit, mathematically represented as a unitary operator $U$, and its noisy implementation, mathematically represented as a super-operator $\e$. Our main references for this section are \cite[Chapter 9]{NC00} and \cite{Gil05}.

To this end, we first 
recall that the \emph{fidelity} between two density operators $\rho$ and $\sigma$ on a Hilbert space $\mathcal{H}$
 is defined by $$F(\rho,\sigma)=(\tr \sqrt{\rho^{1/2}\sigma\rho^{1/2}})^2.$$ When $\rho =|\psi\rangle\langle\psi|$ is a pure state, $F(\psi,\sigma)=\langle\psi|\sigma|\psi\rangle$.

The fidelity between density operators can be extended to measure distance between super-operators with the help of \textit{Jamiolkowski isomorphism} that maps each super-operator $\mathcal{E}$ on $\mathcal{H}$ to a density operator $\rho_{\mathcal{E}}=(\mathcal{I}\otimes\mathcal{E})(|\Psi\rangle\langle\Psi|)$ on $\mathcal{H}\otimes\mathcal{H}$, where $\mathcal{I}$ is the identity super-operator on $\mathcal{H}$, $|\Psi\rangle=\frac{1}{\sqrt{d}}\sum_i|ii\rangle$ is the maximally entangled state, $d=\dim\mathcal{H}$, and $\{|i\rangle\}$ is an orthonormal basis of $\mathcal{H}$. 
The \emph{Jamiolkowski fidelity} between super-operators $\mathcal{E}$ and $\mathcal{F}$ is defined as
$$ 
F_J(\mathcal{E},\mathcal{F})=F(\rho_\mathcal{E},\rho_\mathcal{F}).$$
In the special case when $\mathcal{F} = \{U\}$ is a unitary operation, which is exactly what we are concerned with in this paper, we slightly abuse the notation to write $F_J(\mathcal{E},U)$ instead of $F_J(\mathcal{E},\mathcal{F})$.

Note that there are a wide range of distance measures for density operators and super-operators presented in the literature~\cite{Gil05}, each having applications in different quantum information tasks. Our design decision to choose Jamiolkowski fidelity to characterise the approximate equivalence of an ideal circuit and its noisy implementation is based on the following observations.

{\bf Physical interpretation.}
The Jamiolkowski fidelity $F_J$ can serve as the average-case error measure in computation of a function or sampling distribution \cite{Gil05}, which is particularly useful in quantum simulation and quantum machine learning. To be specific, for super-operator $\e$ and unitary operator $U$, we have
$$\int d\psi F(\mathcal{E}(\psi), U|\psi\rangle) = \frac{d\cdot F_J(\mathcal{E},U)+1}{d+1},$$
where the integral is over the Haar measure on $\mathcal{H}$.
In other words, $F_J(\mathcal{E},U)$ characterises the \emph{average fidelity} of the output $\mathcal{E}(\psi)$ of the noisy circuit $\e$ and that of the ideal one $U|\psi\rangle$ , when the input pure state $|\psi\rangle$ is chosen at random.

{\bf Nice properties.} It is easy to prove that the Jamiolkowski fidelity enjoys the following properties (see \cite{Gil05}):
\begin{enumerate}
	\item \textit{Stability}: $F_J(\mathcal{E}\otimes\mathcal{I}, \mathcal{F}\otimes\mathcal{I})= F_J(\mathcal{E}, \mathcal{F})$, where $\mathcal{I}$ stands for the identity operation on an arbitrary ancillary system; 
	\item \textit{Chaining}: $C_J(\mathcal{E}_1\circ \mathcal{E}_2, U_1\circ U_2)\leq C_J(\mathcal{E}_1, U_1)+C_J(\mathcal{E}_2, U_2)$,
	where $C_J(\mathcal{E}, \mathcal{F}) = \sqrt{1-F_J(\mathcal{E}, \mathcal{F})}$ is a metric between super-operators induced by $F_J$.
\end{enumerate} 
In the scenario of approximate equivalence checking, these properties guarantee that the error scales at most linearly when smaller noisy quantum circuits are concatenated into a larger one.

{\bf Ease of calculation.} Given an ideal circuit $U$ and a noisy implementation $\mathcal{E} = \{E_i\}$, we have
\begin{align*}
		F_J(\mathcal{E},U)&
		= \sum_i \bra{\Psi} I\otimes U^{\dagger} E_i\ket{\Psi}\bra{\Psi} I\otimes E_i^\dag U \ket{\Psi}\\
		&=\sum_i |\bra{\Psi} I\otimes U^{\dagger} E_i\ket{\Psi}|^2=\frac{1}{d^2}\sum_{i}{|\tr(U^{\dagger}E_i)|^2}.
\end{align*}
Thus the calculation of Jamiolkowski fidelity $F_J(\e, U)$ boils down to computing the traces of some matrices. As we are going to show in the following sections, this can be done quite easily by employing the recently proposed Tensor Decision Diagram (TDD) techniques~\cite{tdd}.

To conclude this section, we propose the notion of approximate equivalence between quantum circuits.
\begin{defn} 
Let $\epsilon\in [0,1]$, and $\mathcal{C}_1, \mathcal{C}_2$ be two (noisy) quantum circuits represented by super-operators $\mathcal{E}_1, \mathcal{E}_2$, respectively. Then $\mathcal{C}_1$ and $\mathcal{C}_2$ are $\epsilon$-equivalent, written $\mathcal{C}_1\approx_{\epsilon}\mathcal{C}_2$, if $F_J(\mathcal{E}_1,\mathcal{E}_2)>1-\epsilon$.
\end{defn}
The problem of approximate equivalence checking of noisy quantum circuits can then be formalised as follows:

\begin{prob}
	Let $\epsilon\in [0,1]$ be an error threshold.
	Given (the classical description of) an ideal circuit $\mathcal{C}$ and a noisy implementation $\mathcal{N}$, determine if $\mathcal{C}\approx_{\epsilon}\mathcal{N}$.  
\end{prob}

\section{Algorithms for Approximate Equivalence Checking}\label{Sec_Algo}

In the previous section, we formalise the problem of approximate equivalence checking for quantum circuits. This section is devoted to two algorithms solving this problem in different situations.

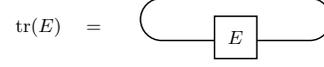
\begin{figure}
\centering
\scalebox{0.8}{
\begin{tikzpicture}
  	\node at (-3, 0.25) [anchor=north]{\begin{tikzcd} 
 			 {\tr(E)} \quad = 
	\end{tikzcd}};
	
\node (start) at (0,0) [draw, terminal,minimum width=90, minimum height=20,line width =0.7]{};
\node (process) at (0,-0.3) [draw, process, minimum width=20, minimum height=20,fill=white!30,line width =0.7] {$E$};
\end{tikzpicture}
}
\caption{The tensor network of computing the trace of the functionality matrix $E$ of a (noiseless or noisy) circuit.}
\label{fig:trace}
\end{figure}

Before diving into the details, we first note that when regarding a quantum circuit as a tensor network, the trace of the functionality matrix represented by the circuit can be calculated by connecting the input qubits with the corresponding output ones and contracting the obtained network. Fig.~\ref{fig:trace} shows a general scheme of computing the trace of the functionality matrix $E$ of a (noiseless or noisy) circuit.


\subsection{Algorithm I: Calculate Traces Individually} 

Recall that $F_J(\mathcal{E},U) = \sum_{i}{|\tr(U^{\dagger}E_i)|^2}/d^2$ where $\e = \{E_i\}$. 
Our first algorithm calculates $\tr(U^{\dagger}E_i)$ for each $E_i$ and adds up the squares of their norms.
Specifically, suppose there are $m$ noisy gates $\n_1, \cdots, \n_m$ in the circuit and  $\n_k=\left\{N^k_1, \cdots, N^k_{n_k}\right\}$ for each $k$. Then every choice of $j_k \in [1, n_k]$, $1\leq k\leq m$, determines a Kraus operator $E_i$ of $\e$ by replacing each $\n_k$ in the noisy circuit with the non-unitary gate $N^k_{j_k}$. Note that the total number of different choices is $\prod_k n_k$. Furthermore, as the ideal circuit $U$ represents a unitary operator, we obtain a circuit representing $U^\dag$ by replacing each gate in $U$ with its Hermitian conjugate and reversing the order of gates. Then $\tr(U^{\dagger}E_i)$ can be computed by concatenating the circuits of $E_i$ and $U^{\dagger}$ (the order is irrelevant), connecting the corresponding input and output qubits, and then contracting the obtained tensor network.



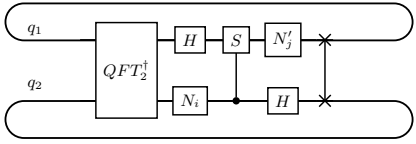
\begin{figure}
\centering

\scalebox{0.7}{
\begin{tikzpicture}
\node  at (0.1,-0.205) [draw, terminal,minimum width=220, minimum height=20,line width =1]{};
\node  at (0.1,-2.06) [draw, terminal,minimum width=220, minimum height=20,line width =1]{};
 minimum height=50,line width =1]{};
\node at (0,0) [anchor=north]{
\begin{tikzcd}[column sep=0.3cm]
&\gate[2]{QFT_2^{\dagger}}&\gate{H}&\gate{S}&\gate{N_j^{\prime}}&\swap{1}\\
&&\gate{N_i}&\ctrl{-1}&\gate{H}&\targX{}
\end{tikzcd}};
\node at (-3.2, -0.15) [anchor=north]{$q_1$};
\node at (-3.2, -1.2) [anchor=north]{$q_2$};
\end{tikzpicture}
}



\caption{Tensor networks to be contracted in Example~\ref{exam:algo1} using Alg. I.
}
\label{exp-for-method1} 
\end{figure}

\begin{exam}\label{exam:algo1}
Consider the noisy circuit $\e$ shown in Fig.~\ref{exp-for-noisy-circuit} which implements the 2-qubit quantum Fourier transform QFT$_2$. 
Suppose the noisy gate $\n=\{N_1, N_2\}$ represents a bit flip, with $N_1=\sqrt{p}I$ and $N_2=\sqrt{1-p}X$, and ${\n}'=\{N'_1, N'_2\}$ represents a phase flip, with $N'_1=\sqrt{p}I$ and $N'_2=\sqrt{1-p}Z$. Then, we have four tensor networks that need to be contracted, all having form shown in Fig.~\ref{exp-for-method1}. 
Contracting these tensor networks gives 
$\tr(U^{\dagger}E_{1,1})=4p$, and $\tr(U^{\dagger}E_{1,2})=\tr(U^{\dagger}E_{2,1})=\tr(U^{\dagger}E_{2,2})=0$, where $U =$ QFT$_2$, $E_{i,j}$ is the functionality matrix of the circuit in Fig.~\ref{exp-for-noisy-circuit} with $\n$ being replaced by $N_i$ and $\n'$ replaced by $N_j'$. Thus $F_J(\e, U)= p^2$.
\end{exam}


Note that the number of circuits to be contracted increases exponentially w.r.t. the number of noisy gates in the circuit. This method will be efficient only when there are very few noisy gates. However, we note that distributed computing techniques can help in contracting these circuits in parallel, since they are completely independent. Furthermore, for the purpose of approximate equivalence checking, it is often sufficient to calculate only a small part of these trace terms, as an affirmative answer is obtained once a partial sum is greater than $1-\epsilon$. Consider the above example again. Suppose $p=0.95$ and our aim is to check if $\mathcal{E}\approx_{0.1} \mathcal{U}$. Clearly, computing $\tr(U^\dag E_{1,1})$
already suffices as $F_J(\e, U) \geq (4p)^2/{4^2} = 0.9025 > 0.9 = 1- 0.1$. 
In practice, this can partially alleviate the problem caused by the exponential number of trace calculations.


%
%

\subsection{Algorithm II: Calculate Traces Collectively} 
We now introduce the second algorithm for approximate equivalence checking, which calculates $\sum_{i}{|\tr(U^{\dagger}E_i)|^2}$ collectively by contracting a single (in contrast to the exponentially many in Alg.~I) tensor network, although it is twice bigger than the tensor networks contracted in Alg.~I.

\begin{figure}
\centering
\scalebox{0.7}{
\begin{tikzpicture}
\node  at (0.1,-0.13) [draw, terminal,minimum width=220, minimum height=20,line width =1]{};
\node  at (0.1,-0.46) [draw, terminal,minimum width=240, minimum height=50,line width =1]{};
\node  at (0.1,-3.59) [draw, terminal,minimum width=220, minimum height=20,line width =1]{};
\node  at (0.1,-3.18) [draw, terminal,minimum width=240, minimum height=51,line width =1]{};
\node at (0,0) [anchor=north]{
\begin{tikzcd}[row sep=0.2cm, column sep=0.2cm]
&\gate[2]{QFT_2^T}_{q_1^{\prime}} &\gate{H} &\qw &\gate{S^*}&\qw&\gate[3,label style={yshift=0.3cm}]{M_{\mathcal{N^{\prime}}}}&\swap{1}\\
&\qw &\qw &\gate[3,label style={yshift=0.3cm}]{M_{\mathcal{N}}}&\ctrl{-1}&\gate{H}&\qw&\targX{}\\
&\gate[2][1.5cm]{QFT_2^{\dagger}}&\gate{H}& \linethrough&\gate{S}&\qw&\qw&\swap{1}\\
& &\qw                       &&\ctrl{-1}&\gate{H}&\qw&\targX{}\\
\end{tikzcd}};
\node at (-3.4, -1.9) [anchor=north]{$q_1$};
\node at (-3.4, -2.85) [anchor=north]{$q_2$};
\node at (-3.4, 0.03) [anchor=north]{$q_1^{\prime}$};
\node at (-3.4, -0.85) [anchor=north]{$q_2^{\prime}$};
\draw[dashed] (-0.7, -2.28) -- (0.3, -2.28);
\draw[dashed] (2, -1.34) -- (3, -1.34);
\end{tikzpicture}
}
\caption{The tensor network to be contracted in Example~\ref{exam:algo2} using Alg. II.}
\label{exp-for-method2}
\end{figure}
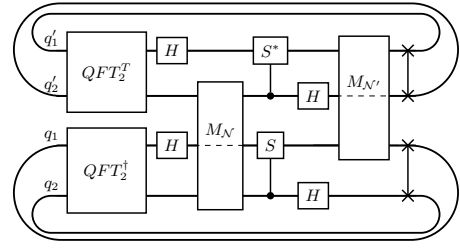

To see how it works, we note that
\begin{equation*}
\begin{aligned}
\sum_{i}{|\tr(U^{\dagger}E_i)|^2} 
&=\sum_{i}{\tr(U^{\dagger}E_i)\cdot \tr(U^{T}E_i^{*})}\\
&=\sum_{i}{\tr(U^{\dagger}E_i\otimes U^{T}E_{i}^{*})} =\tr\big((U^{\dagger}\otimes U^{T})M_{\mathcal{E}}\big),
\end{aligned}
\end{equation*}
where $U^T$ is the transpose of $U$ and $E_i^*$ the complex conjugate of $E_i$,  both with respect to a fixed orthonormal basis of the state space, and for each super-operator $\e = \{E_i\}$, 
$M_{\mathcal{E}}:=\sum_{i}{E_i\otimes E_i^{*}}$ is the \emph{matrix representation} of $\e$~\cite{ernst1987principles}. In other words, the quantity $\sum_{i}{|\tr(U^{\dagger}E_i)|^2}$ as a whole can be computed by the following steps: 
\begin{enumerate}
	\item  introduce for each qubit $q_j$ in the ideal circuit an auxiliary one $q'_j$, and do the same for the noisy circuit;
	\item for each (unitary) gate $V$ acting on $q_{j_1}, \ldots, q_{j_\ell}$ in the ideal circuit, add the complex conjugate gate $V^*$  acting on $q'_{j_1}, \ldots, q'_{j_\ell}$. Denote by $U_E$ the obtained circuit. Note also that $U_E$ consists of two completely separated circuits: the original one $U$ and the newly added one $U^*$;
	\item for each unitary gate in the noisy circuit, do the same as for the ideal circuit; for each noisy gate $\n = \{N_k\}$ acting on $q_{j_1}, \ldots, q_{j_\ell}$, replace it with the corresponding $M_{\mathcal{N}}=\sum_{k}{N_k\otimes N_k^{*}}$, operating on $q_{j_1}, \ldots, q_{j_\ell},q'_{j_1}, \ldots, q'_{j_\ell}$. Denote by $\e_E$ the obtained circuit;
	\item concatenate $U^\dag_E$ and $\e_E$ (again, the order is irrelevant), connect the corresponding input and output qubits, and contract the obtained tensor network, as in Alg.~I. 
\end{enumerate}
\begin{exam}\label{exam:algo2}
Back to the noisy circuit shown in Fig.~\ref{exp-for-noisy-circuit} implementing 
QFT$_2$. Let $\n$ and $\n'$ be defined as in Example~\ref{exam:algo1}. Alg.~II requires to contract the (single) tensor network in Fig. \ref{exp-for-method2}. Note that two auxiliary qubits $q'_1$ and $q'_2$ are added, and the noisy gate $\mathcal{N}$ applied on $q_2$ is replaced by a two-qubit gate $M_{\mathcal{N}}=p I\otimes I+(1-p)X\otimes X$ applied on $q_2$ and $q_2^{\prime}$. Similarly, $\n'$ is replaced by $M_{\mathcal{N}'}=p I\otimes I+(1-p)Z\otimes Z$.
Contracting this tensor network then gives the quantity $16p^2$ directly, and the desired Jamiolkowski fidelity $p^2$, coinciding with Example~\ref{exam:algo1}.

\end{exam}


Compared to Alg.~I, the benefit of this new one is obvious: we have only to contract one single tensor network instead of exponentially many. When there are a large number of noisy gates, which is always the case in actual quantum devices since every gate suffers some degree of noise,
this approach will be definitely more efficient. However, the cost we pay here is that the number of qubits doubles, and the circuit to be contracted is more complicated. This will offset the benefit brought by the single tensor network when the number of noisy gates is quite small. This time-space trade off will be clearly demonstrated with experimental results shown in  Sec.~\ref{Sec_experiments}. 




\subsection{Data Structure and Optimisation}\label{sec:opt}
Our algorithms for computing $F_J(\e, U)$ presented above 
are both based on tensor network contraction. To implement these algorithms, we have to find an efficient data structure to help represent and manipulate tensor networks. To this end, we employ
the recently proposed Tensor Decision Diagram (TDD)~\cite{tdd}.
This software package represents a tensor as a decision diagram which optimises the memory consumption. Note that the final TDD of the contracted tensor network in our algorithms has only one node and the weight of its incoming edge is exactly the trace needed. Our algorithms leave more room for optimisation techniques in contracting tensor networks. We briefly discuss four of them in the following.




{\bf Tree decomposition.} We have employed the tree decomposition algorithm proposed in~\cite{Treedecom} to determine the contraction order for tensor networks. A tree decomposition is a mapping of a graph into a tree, while in our circumstances, the graph is the corresponding line graph of the tensor network and each tree node consists of indices of the tensor network. An optimal contraction order can be obtained in a standard way from a tree decomposition \cite{Treedecom}. 




{\bf Computed table.}
In the TDD package \cite{tdd},  a computed table is used to store all nodes of the decision diagrams generated in each trace calculation process. Recall that Alg.~I computes traces for many miter-like tensor networks with the same structure. It is therefore natural to keep only one computed table in the whole process and maximally reuse the computed results. This idea was first introduced in \cite{li2020eliminating} in the simulation of noisy quantum circuits, but implemented with different (more involved) techniques.

{\bf Local optimisations.} Note that a simple optimisation technique for tensor network contraction is to eliminate adjacent gates whose product is the identity operator, for example, a pair of mutually inverse gates~\cite{Markov, burgholzer2020advanced}. This is particularly useful in approximate equivalence checking because, in many cases, the noisy circuit shares most of the unitary gates with the ideal one except for a few noisy gates. Furthermore, as the corresponding input and output qubits are connected to compute the trace, this technique can be used on both ends of the composed circuit as well. 

{\bf SWAP elimination.} Note that in quantum Fourier transform, a series of SWAP gates are used to reverse the order of output qubits. For this type of circuits, we may simply omit these SWAP gates but instead connect the input qubits to the corresponding outputs determined by the SWAP gates. Obviously this does not change the computed trace value. 

\begin{figure}
\centering
\subfigure[]{
\scalebox{0.7}{
\begin{tikzpicture}
\node  at (0.1,-0.18) [draw, terminal,minimum width=220, minimum height=20,line width =1]{};
\node  at (0.1,-1.97) [draw, terminal,minimum width=220, minimum height=20,line width =1]{};
 minimum height=50,line width =1]{};
\node at (0,0) [anchor=north]{
\begin{tikzcd}[column sep=0.3cm]
&\swap{1}&\qw&\gate{S^{\dagger}}&\gate{H}&\gate{H}&\gate{S}&\gate{N_j^{\prime}}&\swap{1}\\
&\targX{}&\gate{H}&\ctrl{-1}&\qw&\gate{N_i}&\ctrl{-1}&\gate{H}&\targX{}
\end{tikzcd}};
\node at (-3.2, -0.15) [anchor=north]{$q_1$};
\node at (-3.2, -1.2) [anchor=north]{$q_2$};
\end{tikzpicture}
}}

\subfigure[]{
\scalebox{0.7}{
\begin{tikzpicture}
\node  at (0.1,-0.18) [draw, terminal,minimum width=220, minimum height=20,line width =1]{};
\node  at (0.1,-1.97) [draw, terminal,minimum width=220, minimum height=20,line width =1]{};
 minimum height=50,line width =1]{};
\node at (0,0) [anchor=north]{
\begin{tikzcd}
&\gate{S^{\dagger}}&\qw&\gate{S}&\qw&\gate{N_j^{\prime}}\\
&\ctrl{-1}&\gate{N_i}&\ctrl{-1}&\qw&\qw
\end{tikzcd}};
\node at (-3.2, -0.15) [anchor=north]{$q_1$};
\node at (-3.2, -1.2) [anchor=north]{$q_2$};
\end{tikzpicture}
}}
\caption{(a) The tensor network  that needs to be contracted in approximate checking the noisy circuit in Fig.~\ref{exp-for-noisy-circuit} against the ideal one in Fig.~\ref{exp-for-quantum-circuit},
and  (b) the simplified one using the optimisation techniques in Sec.~\ref{sec:opt}.}
\label{fig:swap elim}
\end{figure}
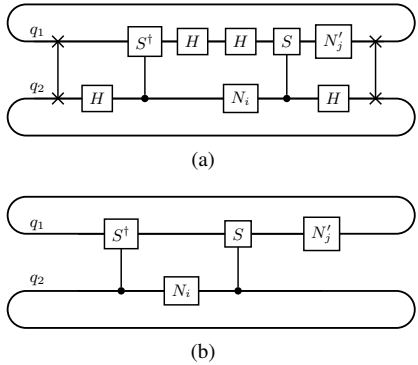

\begin{exam}\label{exam:opt}
	Back to the noisy circuit shown in Fig.~\ref{exp-for-noisy-circuit}, but this time we assume that the ideal circuit is given explicitly in  Fig.~\ref{exp-for-quantum-circuit}. Then according to Alg.~I, we have to contract tensor networks with the form shown in Fig.~\ref{fig:swap elim} (a). Using the optimisation techniques presented above, the two SWAP gate can be eliminated by reconnecting the inputs and outputs and the four $H$ gates can be cancelled locally. Consequently, we only have to contract the much simplified tensor network shown in Fig.~\ref{fig:swap elim} (b). 
\end{exam}

\section{Empirical Evaluations}\label{Sec_experiments}

In this section, we evaluate the effectiveness of our algorithms by comparing with the Qiskit $process\_fidelity$ method. We also illustrate how our algorithms perform as the number of noises increases. Finally, the utility of the computed table in saving computations for Alg.~I is demonstrated. We implement the algorithms using Python3 and conduct experiment on a laptop with Intel i7-1065G7 CPU and 8GB RAM. {The optimisation technologies of tree decomposition and computed table (cf. Sec.~\ref{sec:opt}) 
are incorporated. However, in order to make a fair comparison with Qiskit's corresponding method, local optimisations and SWAP elimination are excluded.}



\subsection{Scalability With the Number of Qubits}
{\bf Baseline} We  compare the time and memory consumption of our algorithms with the corresponding Qiskit implementation.
To this end, we first transform the ideal circuit and the noisy one into Qiskit classes 
$Operator$ and $SuperOp$, respectively.
Then the method $process\_fidelity$ is called to calculate the fidelity.



{\bf Benchmarks}
We experiment on some well-known algorithms, such as  Bernstein-Vazirani algorithm (bv) \cite{bernstein1997quantum}, Quantum Fourier Transform (qft) \cite{NC00}, Grover algorithm \cite{grover1996fast}, as well as benchmarks appeared in real quantum tasks, such as Quantum Volume (qv) \cite{moll2018quantum}, Modular
Multiplication (7x1mod15) \cite{aleksandrowicz2019qiskit}, and Randomised Benchmarking (rb) \cite{knill2008randomized}. All of our benchmarks are from \cite{li2020eliminating}. 

For noisy implementations, we randomly insert some depolarisation noises (see Example~\ref{flips}) to model the realistic errors that occur in real NISQ devices. The numbers of noises range from 1 to 14 and the 
probability parameter of the noisy gate is set to be 0.001 (i.e., $p = 0.999$ in Example~\ref{flips}), representing the state-of-the-art design technology \cite{tannu2019not}. It is worth noting that the selection of $p$ does not affect the performance of the algorithms.


{\bf Results}
Table \ref{experiment-results} summarises the experimental results of the Qiskit $process\_fidelity$ method and our algorithms. In the experiment, the time-out and memory bound are set as 3600 seconds and 8GB, respectively. Note that in Qiskit class $SuperOp$, a super-operator in an $n$-qubit system is stored as a $2^{2n}\times 2^{2n}$ complex matrix, which is extremely space-consuming. For example, at least 64GB memory space is needed to describe an 8-qubit super-operator using the data type $complex128$. 
Consequently, the baseline algorithm can only process circuits with at most 7 qubits on our computer with 8GB RAM.
Even for the two 7-qubit circuits $qft7$ and $qv\_n7d5$, $process\_fidelity$ fails due to memory overflow.

From Table \ref{experiment-results} we can see that, when the number of qubits is small ($\leq 5$), the time consumptions of Qiskit and Alg. II  are very close, but when the number of qubits becomes 6, Alg. II runs faster than Qiskit in several orders of magnitude. The table further shows that our algorithms scale well when the number of qubits increases from 6 to 16. This is partially due to that 
we adopt the decision diagram representation, where the minimum number of complex numbers needed to store for a TDD with $m$ nodes is only $2m-1$. 


From the table we also observe that, when the number of noises is small, Alg.~I could be more efficient than Alg.~II, but when the number of noises becomes greater this will be changed. The following subsection illustrates this in more details.

\begin{table}[]
\centering
\caption{Experiment results}
\scalebox{0.8}{
\begin{tabular}{|l|r|r|r|r|r|r|r|r|}
\hline
\multicolumn{1}{|l|}{\multirow{2}{*}{Circuit}} & \multicolumn{1}{c|}{\multirow{2}{*}{$n$}} & \multicolumn{1}{c|}{\multirow{2}{*}{$|G|$}} & \multicolumn{1}{c|}{\multirow{2}{*}{$k$}} & \multicolumn{1}{c|}{Qiskit} & \multicolumn{2}{c|}{Alg.II}                            & \multicolumn{2}{c|}{Alg. I}                            \\ \cline{5-9} 
\multicolumn{1}{|c|}{}                            & \multicolumn{1}{c|}{}                   & \multicolumn{1}{c|}{}                     & \multicolumn{1}{c|}{}                   & \multicolumn{1}{c|}{time (s)}   & \multicolumn{1}{c|}{time (s)} & \multicolumn{1}{c|}{nodes} & \multicolumn{1}{c|}{time (s)} & \multicolumn{1}{c|}{nodes} \\ \hline
rb       & 2  & 7   & 6  & 0.03   & 0.10   & 16    & 24.34  & 6    \\ \hline
qft2     & 2  & 7   & 2  & 0.05   & 0.05   & 12    & 0.14   & 8    \\ \hline
grover   & 3  & 96  & 4  & 0.15   & 3.56   & 283   & 23.58  & 38   \\ \hline
qft3     & 3  & 18  & 7  & 0.16   & 0.53   & 116   & 230.82 & 13   \\ \hline
qv\_n3d5 & 3  & 50  & 2  & 0.21   & 0.91   & 32    & 1.55   & 16   \\ \hline
bv4      & 4  & 11  & 7  & 0.24   & 0.17   & 30    & 210.48 & 8    \\ \hline
7x1mod15 & 5  & 14  & 3  & 4.29   & 0.18   & 14    & 1.39   & 10   \\ \hline
bv5      & 5  & 14  & 6  & 5.13   & 0.26   & 22    & 64.01  & 8    \\ \hline
qft5     & 5  & 55  & 3  & 31.90  & 2.44   & 270   & 4.07   & 62   \\ \hline
qv\_n5d5 & 5  & 100 & 3  & 20.12  & 10.82  & 272   & 33.82  & 256  \\ \hline
bv6      & 6  & 17  & 14 & 258.15 & 0.48   & 30    &  TO      &  TO    \\ \hline
qv\_n6d5 & 6  & 150 & 1  & 1158.10& 18.54  & 256   & 8.94   & 256  \\ \hline
qft7     & 7  & 112 & 6  &  MO      & 33.72  & 1092  & 391.74 & 345  \\ \hline
qv\_n7d5 & 7  & 150 & 2  &   MO     & 72.17  & 928   & 151.04 & 1024 \\ \hline
bv9      & 9  & 26  & 6  &  MO      & 0.52   & 46    & 117.18 & 8    \\ \hline
qv\_n9d5 & 9  & 200 & 3  &  MO      & 82.98  & 962   & 361.29 & 1673 \\ \hline
qft9    & 9 & 189 & 2  &  MO      & 131.41 & 2820 & 27.32  & 1216 \\ \hline
qft10    & 10 & 235 & 2  &  MO      & 595.30 & 15753 & 194.08  & 3012 \\ \hline
bv13     & 13 & 38  & 4  &  MO      & 0.81   & 16    & 11.48  & 8    \\ \hline
bv14     & 14 & 41  & 4  &  MO      & 0.86   & 14    & 14.31  & 8    \\ \hline
bv16     & 16 & 47  & 9  &  MO      & 1.61   & 22    &  TO      &  TO    \\ \hline
\end{tabular}}

\begin{tablenotes}
\item[1]* $n$, $|G|$, $k$ are, respectively, the numbers of qubits, gates, noises. The  `nodes' columns record the maximum numbers of the nodes of the TDDs constructed in the calculation process.
\item[2]* `TO' stands for time out, `MO' stands for out of memory.
\end{tablenotes}

\label{experiment-results}
\end{table}

\subsection{Scalability With the Number of Noises}
In this subsection we examine the performance of Algs.~I and II when the number of noises increases.

{\bf Benchmarks} 
We select Bernstein-Vazirani algorithm and Quantum Fourier Transform as our benchmark circuits of this part. The numbers of qubits in these circuits range from 3 to 5. 

\begin{figure}
    \centering
    \includegraphics[width=0.3\textwidth]{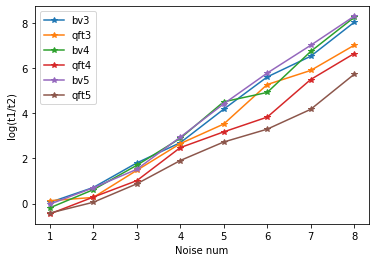}
    \caption{The comparison of the two algorithms as the number of noisy gates increases. The horizontal axis is the number of noises and the vertical axis is $\log(t_1/t_2)$, where $t_1$ and $t_2$ are the time for Alg.~I and Alg.~II respectively.}
    \label{noisy_num_compare}
\end{figure}

{\bf Results}
The results are summarised in Fig.~\ref{noisy_num_compare}, where the polyline `bv3' represents the logarithm of the time consumption ratio $\log(t_1/t_2)$ of the two algorithms on the 3-qubit Bernstein-Vazirani algorithm, and the meaning of other polylines is similar. From
the figure, we can see that when there is only one noise, for most of the circuits, we have $\log(t_1/t_2) < 0$, which means $t_2 > t_1$, i.e., the time consumption of Alg.~II is bigger than that of Alg.~I. But when the number of noises increases, $\log(t_1/t_2)$ increases linearly, meaning that the running time of Alg.~I increases exponentially compared to that of Alg.~II. 
This suggests that Alg.~II is more suitable when the circuit contains many noisy gates. 

\subsection{The Utility of the Computed Table}


To test the optimisation technique of introducing computed tables (see Sec.~\ref{sec:opt}), we check the noisy implementation of Bernstein-Vazirani algorithm using Alg.~I, with the numbers of qubits and noises ranging from 3 to 5 and from 1 to 8, respectively. Table \ref{Use-com-table} shows the experimental results where the columns  `Ori.' and `Opt.' record, respectively, the running time of Alg.~I without and with the shared computed table. It can be observed that an average of 72\%, 62\%, and 65\% of time consumption can be saved for these circuits by reusing the computed table, which is comparable to the result of \cite{li2020eliminating}. 

\begin{table}[]
\caption{Utility of the computed table}
\centering
\scalebox{0.7}{
\begin{tabular}{|c|l|l|l|l|l|l|l|l|l|}
\hline
\multirow{2}{*}{\begin{tabular}[c]{@{}c@{}}Noise\\ num\end{tabular}} & \multicolumn{3}{c|}{bv3}                                                          & \multicolumn{3}{c|}{bv4}                                                          & \multicolumn{3}{c|}{bv5}                                                          \\ \cline{2-10} 
                                                                     & \multicolumn{1}{c|}{Opt.} & \multicolumn{1}{c|}{Ori.} & \multicolumn{1}{c|}{Rate} & \multicolumn{1}{c|}{Opt.} & \multicolumn{1}{c|}{Ori.} & \multicolumn{1}{c|}{rate} & \multicolumn{1}{c|}{Opt.} & \multicolumn{1}{c|}{Ori.} & \multicolumn{1}{c|}{Rate} \\ \hline
1 & 0.05   & 0.10    & 0.50 & 0.11   & 0.13    & 0.82 & 0.10    & 0.17    & 0.56 \\ \hline
2 & 0.17   & 0.44    & 0.38 & 0.35   & 0.53    & 0.66 & 0.30    & 0.67    & 0.45 \\ \hline
3 & 0.63   & 1.84    & 0.34 & 1.06   & 2.20    & 0.48 & 1.00    & 2.63    & 0.38 \\ \hline
4 & 2.26   & 7.52    & 0.30 & 3.87   & 8.76    & 0.44 & 3.89    & 10.83   & 0.36 \\ \hline
5 & 8.55   & 34.03   & 0.25 & 15.91  & 34.09   & 0.47 & 15.13   & 44.07   & 0.34 \\ \hline
6 & 37.24  & 133.59  & 0.28 & 57.87  & 142.44  & 0.41 & 62.77   & 182.12  & 0.34 \\ \hline
7 & 142.92 & 535.77  & 0.27 & 225.86 & 584.75  & 0.39 & 254.34  & 735.31  & 0.35 \\ \hline
8 & 593.56 & 2119.26 & 0.28 & 913.23 & 2445.28 & 0.37 & 1090.03 & 3051.05 & 0.36 \\ \hline
SUM & 785.39 & 2832.56 & 0.28 & 1218.27 & 3218.18 & 0.38 & 1427.56 & 4026.87 & 0.35 \\ \hline
\end{tabular}}
\begin{tablenotes}
\item[1]* Running time (in seconds) of Alg.~I w/ and w/o computed tables.
\item[2]* The rate is calculated by Opt./Ori.
\end{tablenotes}
\label{Use-com-table}
\end{table}

\section{Conclusion and Future works}\label{Sec_Con}
Although the equivalence checking of quantum (noiseless) circuits has been studied for more than ten years, little attention was paid to noisy circuits. In this paper, we defined the approximate equivalence of noisy quantum circuits and proposed two algorithms to check it. Our algorithms are based on calculating the Jamiolkowski fidelity, which is reduced to calculating the traces of miter-like tensor networks. We implemented our algorithms by using the newly proposed data structure TDD. When compared with the current Qiskit method for calculating the Jamiolkowski fidelity, experiments on various real benchmark circuits show that our algorithms outperform in both efficiency and scalability, especially when the circuits have five or more qubits. When comparing our algorithms, Alg.~II is more efficient when many errors occur in the noisy circuit and Alg.~I works better when errors are rare. In addition, for the purpose of approximate equivalence checking, as we often only need to calculate a small part of the involved trace terms, Alg.~I may be more attractive.

Future work will incorporate more optimisation techniques like local optimisations and SWAP elimination (cf. Sec.~\ref{sec:opt}) in the implementation of our algorithms. We will also consider how to select a small subset of trace terms to efficiently approximate the fidelity computation in Alg.~I.

{
\footnotesize
\bibliographystyle{IEEEtran}
\bibliography{references2}
}

\end{document}